\documentclass{evn2004}
\usepackage{txfonts}
\usepackage{graphicx}
\begin{document}
   \title{A detailed study of the nuclear region of Mrk 273} 

   \author{M. Bondi\inst{1}
          \and
          D. Dallacasa\inst{2} \and M.A. P\'erez Torres\inst{3} \and
	  T.W.B. Muxlow\inst{4}
          }

   \institute{Istituto di Radioastronomia, Via Gobetti 101, I-40129, Bologna,
              Italy
         \and
             Dipartimento di Astronomia, Universit\`a di Bologna, Via
             Ranzani 3, I-40127, Bologna, Italy
	 \and   
             Instituto de Astrofisica de Andalucia, CSIC, Apartado Correos
             3004, 18080 Granada, Spain
	 \and
	     Jodrell Bank Observatory, University of Manchester, Macclesfield,
	     Cheshire, SK119DL, U.K.
             }

   \abstract{We present 5 GHz EVN+MERLIN observations of the nuclear region 
   of the
ultra luminous infrared galaxy Mrk 273. These observations confirms the
detection, on the mas scale, of two resolved component labelled as N and SE
in the literature. We use published VLBA observations at 1.4 GHz, at the
same resolution, to derive spectral index information of component N and SE
and discuss these findings in relation with different hypothesis (compact
starburst or AGN) for the origin of the radio emission.
}

   \maketitle
%

\section{Introduction}

Ultraluminous infrared galaxies (ULIRGs) are the most luminous galaxies
in the local universe with  $L \ge 10^{12}$ L$_\odot$,
comparable with quasars (Sanders et al. \cite{Sand88}).
The bulk of the luminosity of ULIRGs
is thought to be produced by dust, in the inner kpc of the galaxy, heated by a
powerful source of optical-UV continuum. An important question to be
answered is which is the dominant gas-heating mechanism in the ULIRGs:
an active galactic nucleus (AGN) or a massive starburst? Recent developments
using near-IR spectroscopy suggest enhanced star formation in the
majority of the ULIRGs (Genzel et al. \cite{Genz98}) with a significant 
heating from the AGN only in the most luminous objects (Veilleux, Sanders \& 
Kim \cite{VSK99}).
Radio observations can prove to be extremely important
in studying the central region of ULIRGs since they are unaffected by dust
extinction and allow for sub-parsec resolution using VLBI techniques.
The most spectacular evidence of a compact starbust in a ULIRG is indeed
the discovery of a population of bright radio supernovae in the nuclear
region of Arp~220 (Smith et al. \cite{Smit98}) detected at 1.6 GHz.

Mkn~273 is a ULIRG at $z=0.0378$ classified as a Seyfert 2 and/or LINER.
It is a merging galaxy showing a disturbed morphology on the kpc scale.
The nuclear region of Mkn~273 is extremely complex and has been
studied in details in the radio (Cole et al. \cite{Cole99},
Carilli \& Taylor \cite{CT00}, Yates et al. \cite{Yate00}), NIR (Knapen et 
al. \cite{Knap97}),
optical (Mazzarella \& Boroson \cite{MB93}), and X--ray band (Xia et al.
\cite{Xia02}).
Three extended radio
components (N, SE and SW, see Fig. 1 in Yates et al. \cite{Yate00}) 
are detected
within 1 arcsecond in the nucleus of Mkn~273. These are all physically
related components, probably associated with the merging process, and
not the result of chance projection of background sources.

The three components show different morphologies and properties.
Component SW is the weaker one and is barely detected only with MERLIN at 18 
cm. Component N
and SW have bright NIR counterparts (Knapen et al. \cite{Knap97}; 
Majewski et al. \cite{Maje93})
and are redder than component SE, suggesting the presence of strong star
formation in these 2 components, while component SE is not detected in the
NIR. 

Throughout this contribution we will assume $H_0=75$ km s$^{-1}$ Mpc$^{-1}$.
At the distance of Mrk 273, 1 mas corresponds to 0.7 pc. To put this in
perspective with other ULIRGs or starburst galaxies where the origin of the
radio emission has been investigated, Mrk 273 is at about twice the distance
of Arp 220 and 56 times the distance of M 82.

\section{The radio view of the nuclear region in Mrk 273}

In this section we will focus on published results from high resolution
radio imaging of components N and SE in Mrk 273. MERLIN observations at 5 GHz
have resolved component N into two compact regions (we will refer to them as N1
and N2 for the western and eastern components, respectively) embedded in 
diffuse emission while component SE is barely resolved.

\begin{figure*}
   \centering
   \includegraphics[height=10.0cm]{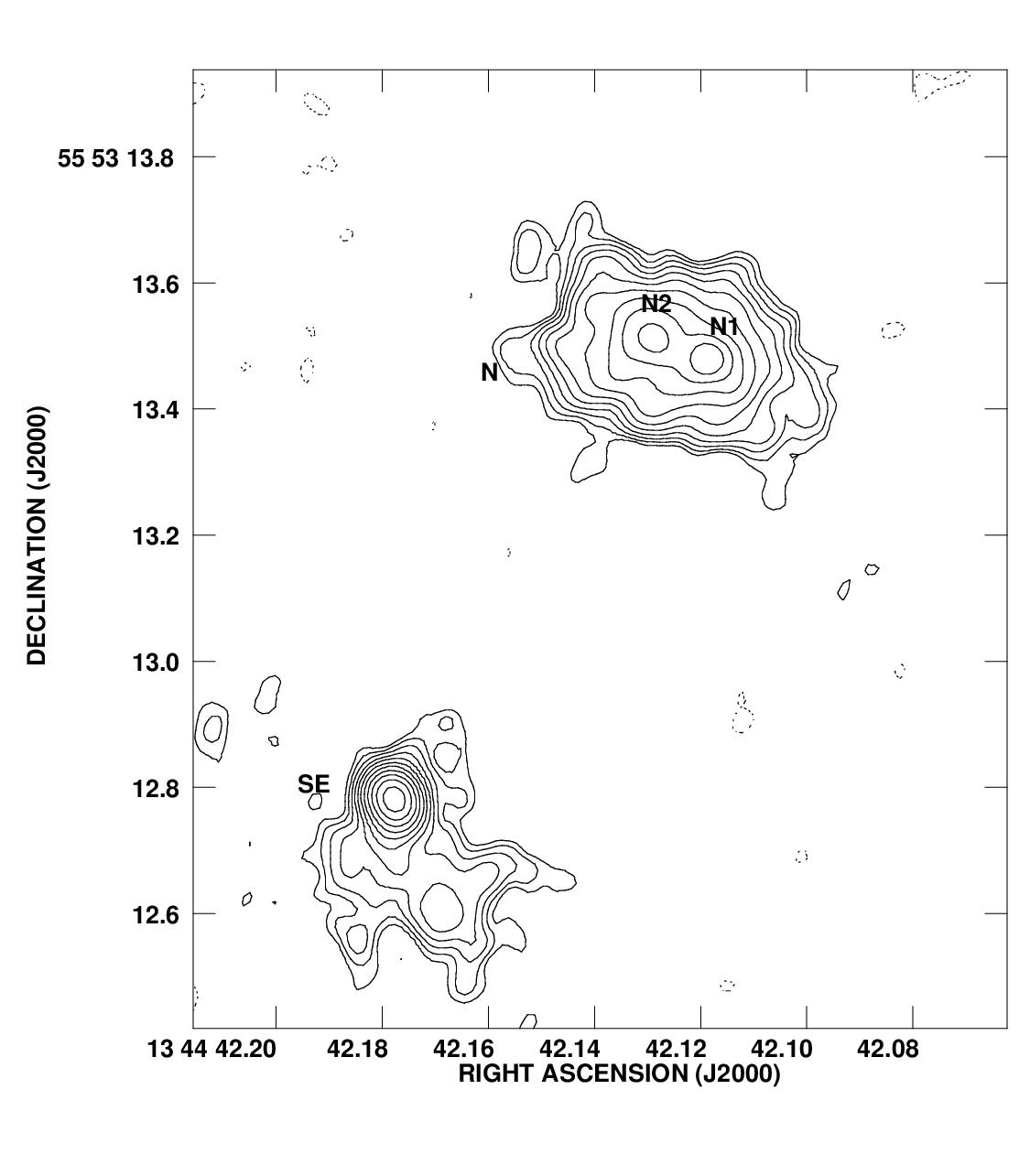}
   \includegraphics[height=10.0cm]{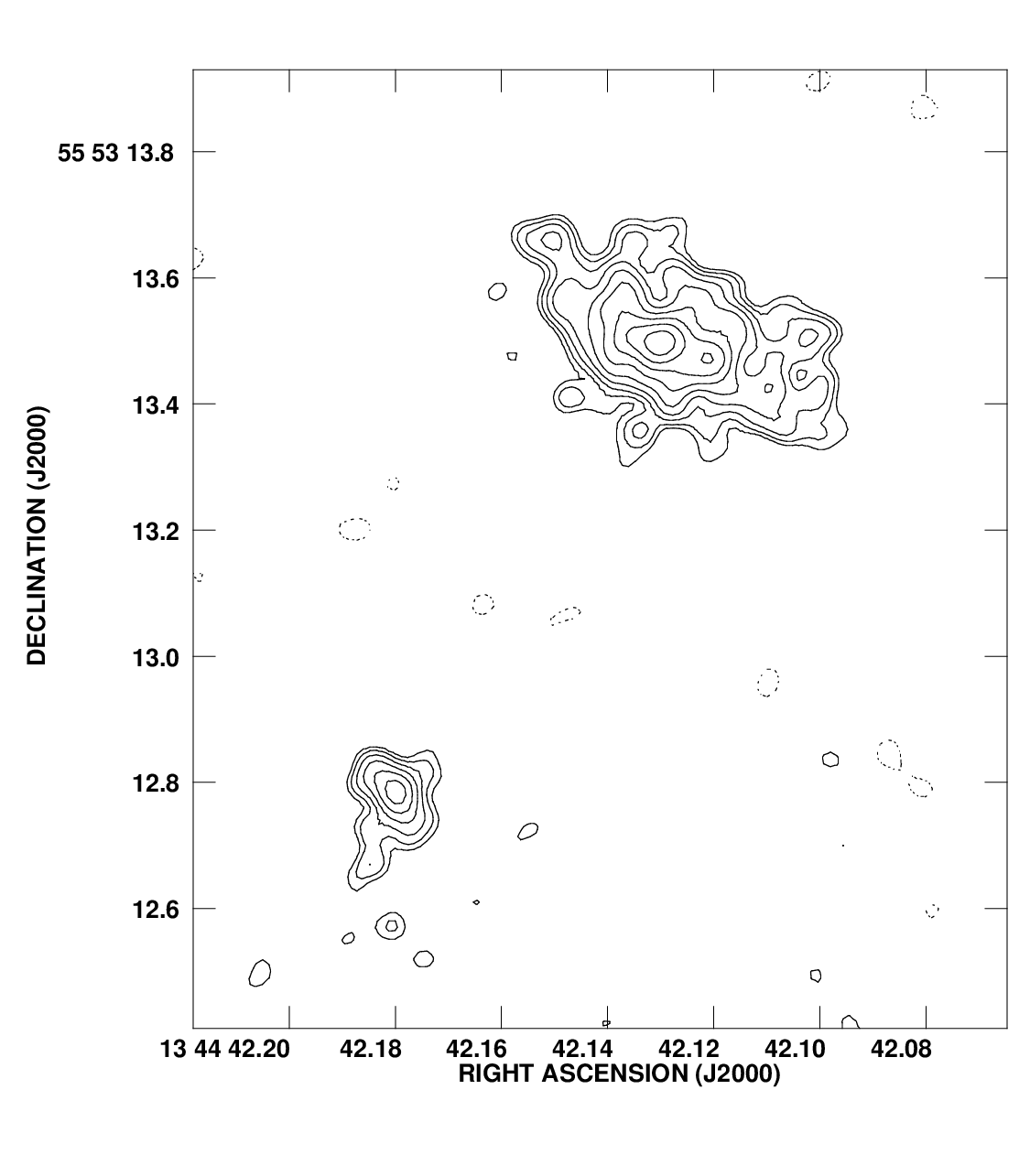}
   \caption{{\it a) Left:} Image of Mrk 273 at 1.37 GHz obtained by the
VLBA+VLA at 50 mas resolution. The contour levels are a geometric
progression in the square root of 2, starting from 0.25 mJy/beam. The peak
surface brightness is 10 mJy/beam and the off-source noise rms is 85 
$\mu$Jy/beam. The labels indicate the two sub-arcsecond scale compact
component N and SE, and the two peaks of emission N1 and N2 inside N.
{\it b) Right:} Image of Mrk 273 at 5 GHz obtained by MERLIN at 50
mas resolution. The contour levels are a geometric
progression in the square root of 2, starting from 0.20 mJy/beam. The peak
surface brightness is 4.2 mJy/beam and the off-source noise rms is 63 
$\mu$Jy/beam.}
\end{figure*}

VLBA+VLA observations at 1.4 GHZ have provided the best images of Mrk 273
obtained so far (Carilli \& Taylor \cite{CT00}).
At a resolution of 50 mas (see fig. 1) two things can be immediately noted
from the comparison of the 1.4 and 5 GHz images: 1) the different spectral
properties of component N1 and N2, N1 is stronger than N2 at 1.4 GHz but the
opposite is true at 5 GHz; 2) the 300 mas ($\sim 210$ pc) long tail of extended
emission detected at 1.4 GHz in the SE component is missing from the 5 GHz
image.

At a resolution of 10 mas (see fig. 2a), 
component N1 is still compact while N2 is resolved in multiple sub-mJy
compact components which form a structure roughly elongated in north south
direction. 
The faint halo of extended emission is punctuated by others weak and compact
components. N1 is the brigtest component ($\simeq 3$ mJy/beam) and is 
possibly identified with a weak AGN (but see below).
Carilli \& Taylor (\cite{CT00}) identify
the compact components in N2 with brightness between 0.5 and 0.9 mJy/beam
($T_b\ge 3\times 10^6$ K, $L_{\rm 1.4GHz}\ge
10^{21}$ W Hz$^{-1}$) with nested supernovae remnants (SNR) and/or luminous 
radio supernovae (SN). To account for the observed luminosity of these compact
features, assuming they are nested SNR, 10 or more of the most
luminous M82-type SNR would be required in regions less than 10 mas
(7 pc) in size, while assuming that the compact sources are SN they would be
extremely luminous, an order of magnitude higher than those observed in M~82
and comparable to the supernovae candidates observed in Arp 220.

The interpretation of the radio morphology of component SE is less
straightforward (see fig. 3a): it shows an elongated structure in N-S about
40 mas long embedded
in a weak halo, consistent with an amorphous jet or a very compact starburst.
The lack of NIR emission in component SE could argue in favour of the AGN
interpretation for this source, but the possibility that component
is still obscured at 2.2 $\mu$m can not be ruled out.

\section{EVN+MERLIN Observations}

In order to investigate the complex nature of Mrk 273 we obtained EVN+MERLIN
observations at 5 GHz. EVN observations were carried out at 512 Mbit/s
sustained bit rate to exploit the large bandwidth capabilities of the EVN,
with an array which included all the european antennas. These were the first
observations at 5 GHz for the resurfaced Lovell telescope.
Mrk~273 was observed in phase-reference mode for a total on-source time of
5.5 hours.
The compact source J1337+550 was observed every 5 minutes as phase reference,
while OQ208 and J1310+322 were used to calibrate the bandpass.
Data reduction was performed using the Astronomical Image Processing System
(AIPS). Standard a priori gain calibration was performed using the measured
gains and system temperatures of each antenna. 
The amplitude calibration was refined using the phase reference source.
EVN and MERLIN data sets were reduced separately and then combined to produce
final images with a resolution of $10\times 10$ arcsec and $50\times 50$
arcsec to be compared with the 1.4 GHz images from Carilli \& Taylor which we
have obtained by the authors. The low resolution image is shown in Fig. 1b,
while high resolution images of component N and SE are given in Fig. 2b \& 3b.

\begin{figure*}
\includegraphics[width=6cm, angle=-90]{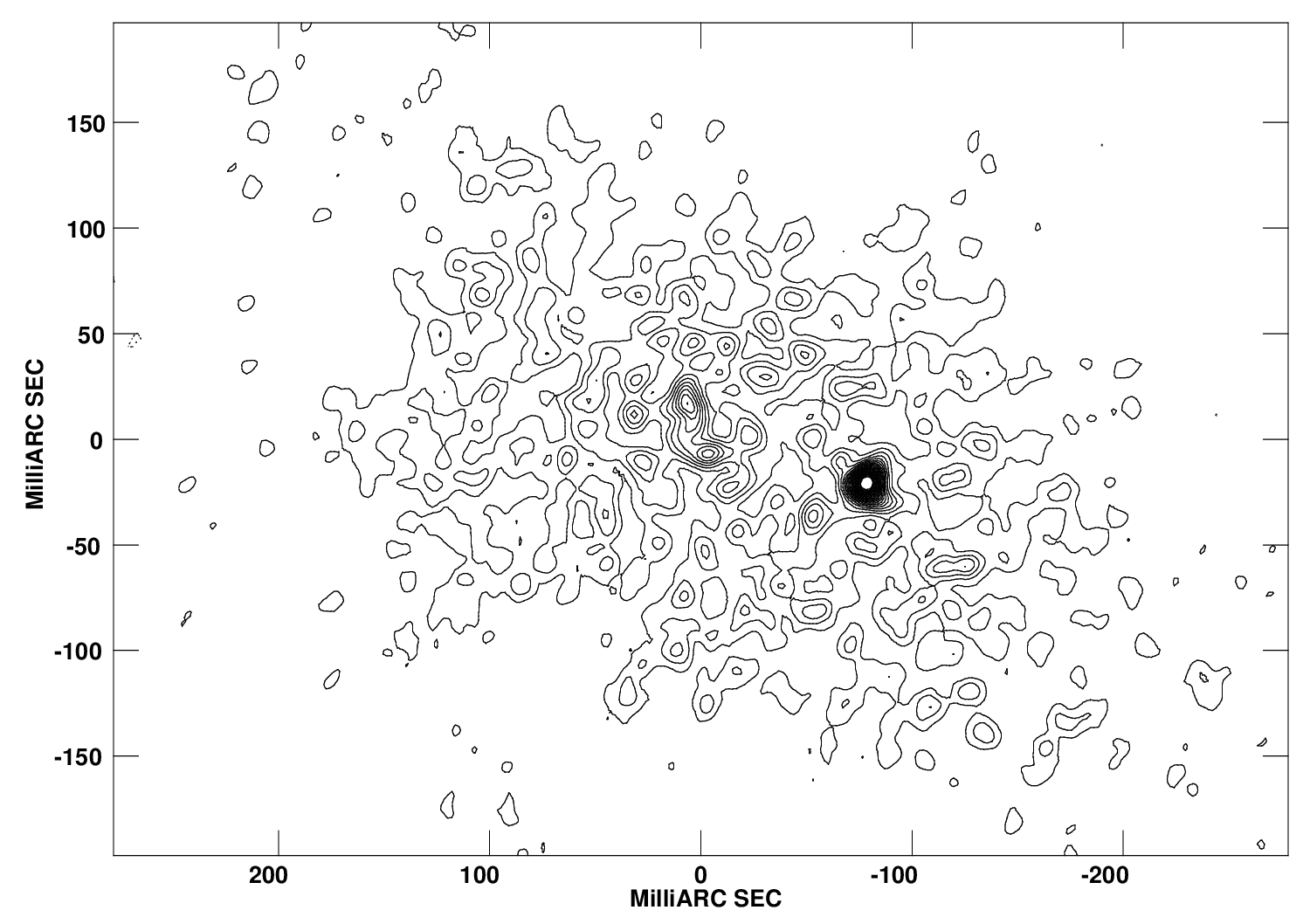}
\includegraphics[width=6cm, angle=-90]{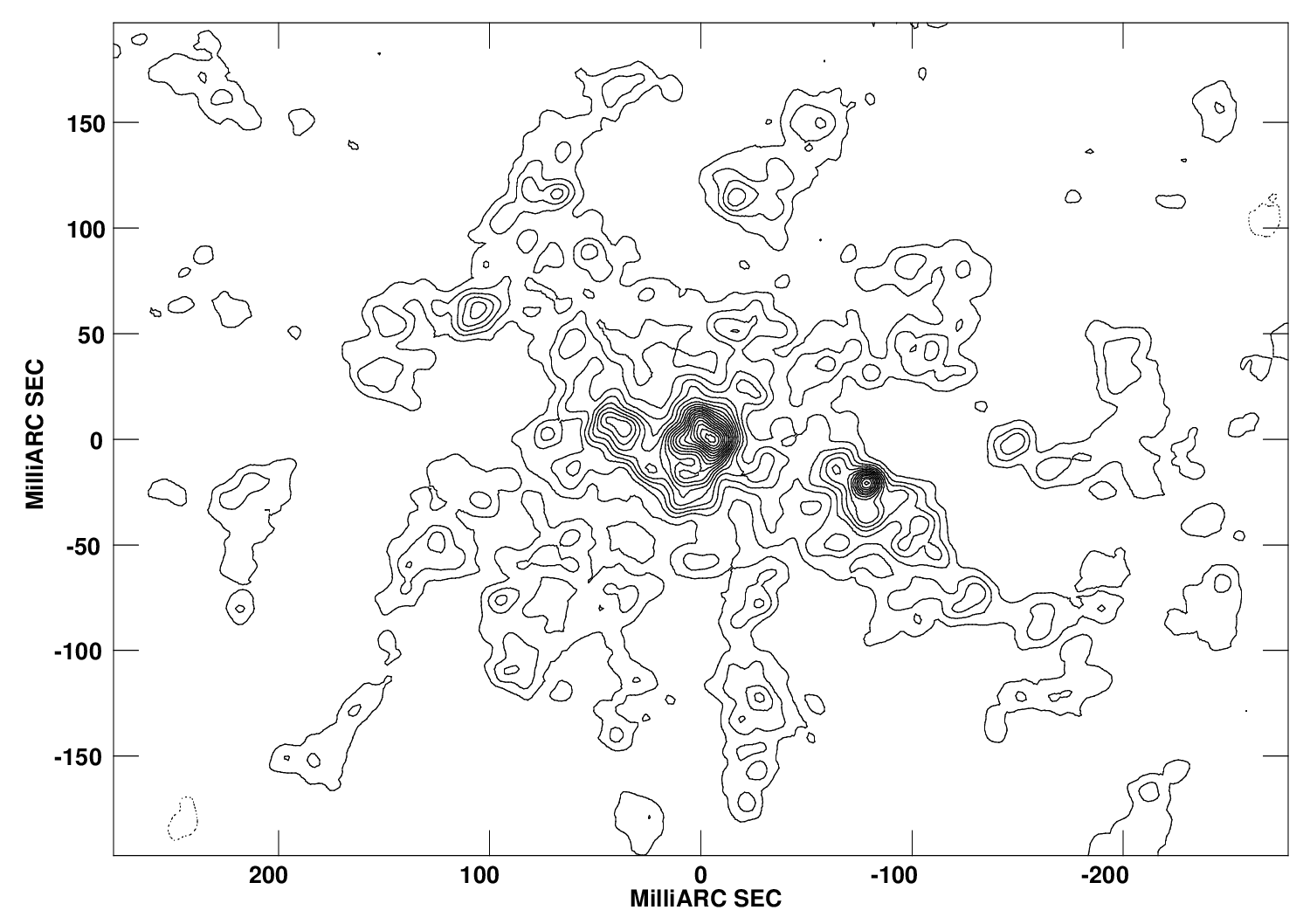}
\caption{{\it a) Left:} VLBA+VLA image of component N at 1.37 GHz at 10 mas
resolution. The contours are $-1, 1, 2, 3, 4, {\ldots} \times 0.1$ mJy/beam.
The peak surface brightness is 3.05 mJy/beam and the off-source noise is 36
$\mu$Jy/beam.
{\it b) Right:} EVN+MERLIN image of component N at 5 GHz at 10 mas resolution.
The contours are $-1, 1, 2, 3, 4, {\ldots} \times 0.04$ mJy/beam.
The peak surface brightness is 0.74 mJy/beam and the off-source noise is 15
$\mu$Jy/beam.
}
\end{figure*}

\section{Results and Discussion}
In this section we present preliminary  results from the EVN+MERLIN
observations at 5 GHz. Using the 1.4 GHz images published in Carilli \& Taylor
(\cite{CT00}), and kindly provided by the authors, we have derived the 
spectral index between 1.4 and 5 GHz and  discuss the origin and spectral 
properties of the emission detected in components N and SE in Mrk 273.
We use the notation $S\propto \nu^{-\alpha}$.

\subsection{Component N in Mrk 273}

The northern source in Mrk 273 consists of two region of compact emission,
coincident with the N1 and N2 components of the lower resolution MERLIN image,
embedded in diffuse emission of roughly the same extension of that detected
at 1.4~GHz by Carilli \& Taylor. It has been supposed, in the
literature, that N1 hosts a weak AGN nucleus and N2 is a very compact region
of massive star formation. This view is supported by two major points:
1) the detection of high excitation IR lines (Genzel et al. \cite{Genz98}) and 
hard X-ray emission (Xia et al. \cite{Xia02}) are evidences of an AGN-like 
nucleus in Mrk 273; 2) the CO emission and NIR peaks are spatially coincident 
with N2 (Downes \& Solomom \cite{DS98}, Knapen et al. \cite{Knap97}).

At 5 GHz, N1 is very weak (0.6 mJy/beam compared with the 3 mJy/beam at 1.4
GHz) yielding to a very steep spectral index $\alpha\simeq 1.2$. 
Fitting the component with elliptical
gaussian and using the total flux density instead of the peak brightness
doesn't change significantly the spectral index.
This is clearly surprising, and would rule out the presence of an AGN nucleus
in N1, unless we consider variability as a possible explanation. 
It is interesting to note that at the full resolution of the EVN array
($\simeq 5$ mas), N1 is slightly resolved and elongated suggesting a core
jet morphology. Deeper observations with an increased uv-coverage are
necessary to resolve the N1 component.

On the other hand, the eastern component, N2, has a very complex morphology
with a flat spectral index $\alpha\simeq 0.15$. The flat spectral index can
be explained as the result of several compact, partly overlapping, components 
with spectra peaking 
at a few GHz and free-free absorption. The radio morphology of this region
is indeed indicative of multiple compact components in a region of about 25
pc, and this is also consistent with the results from CO and NIR emission
observations which indicate this region as the core of an extremely rich
star forming region. Downes \& Salomon (\cite{DS98}) derive an IR luminosity 
of about 
$6\times 10^{11} L_\odot$ generated in a region with a radius of only 120 pc
and a current molecular mass of $1\times 10^9$ $M_\odot$. This means that
the entire molecular core has a radius 5 times that of W51 (a Giant
Molecular Cloud in our galaxy), but with $\sim 3000$ times the molecular mass
and $\sim 10^5$ times the luminosity from OB stars.

The extended emission has a steep spectral index $\alpha\simeq 0.8$.
Following Condon (\cite{Cond92}) the FIR flux ($\log(FIR)=11.6 L_\odot$) can 
be used
to estimate the star formation rate, $\dot m\simeq 40$ $M_\odot$yr$^{-1}$, 
with a corresponding supernovae rate $\nu_{SN}\simeq 2$ yr$^{-1}$. 
These values have been derived using simple scaling law relationship to
estimate starburst characteristics in terms of the star formation rate, the
lower and upper mass limits to the initial mass function, $m_l\sim 5 M_\odot$ 
and $m_u\sim 28 M_\odot$, and the starburst timescale, 
$\Delta t_{SB}\sim 10^8$ yr. Such a
supernovae rate would produce a non-thermal luminosity of $2\times 10^{23}$
W Hz$^{-1}$ at 1.4 GHz. The observed value is $2.2\times 10^{23}$ W
Hz$^{-1}$, in agreement with the predicted one, and so it supports
the hypothesis that the extended emission is produced by elecrons which have
diffused away from the SNR shocks.

\begin{figure*}
\includegraphics[width=6cm]{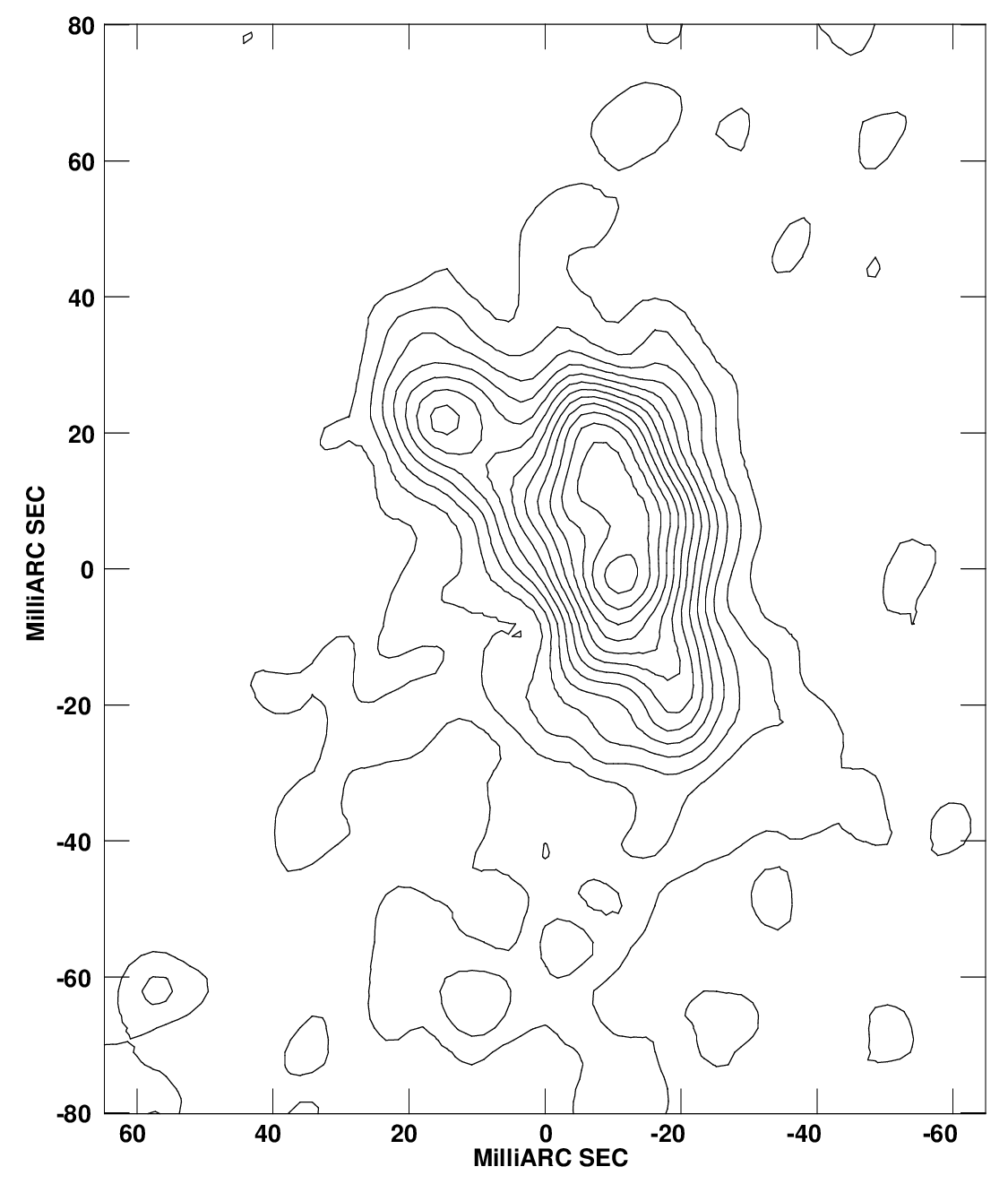}
\includegraphics[width=6cm]{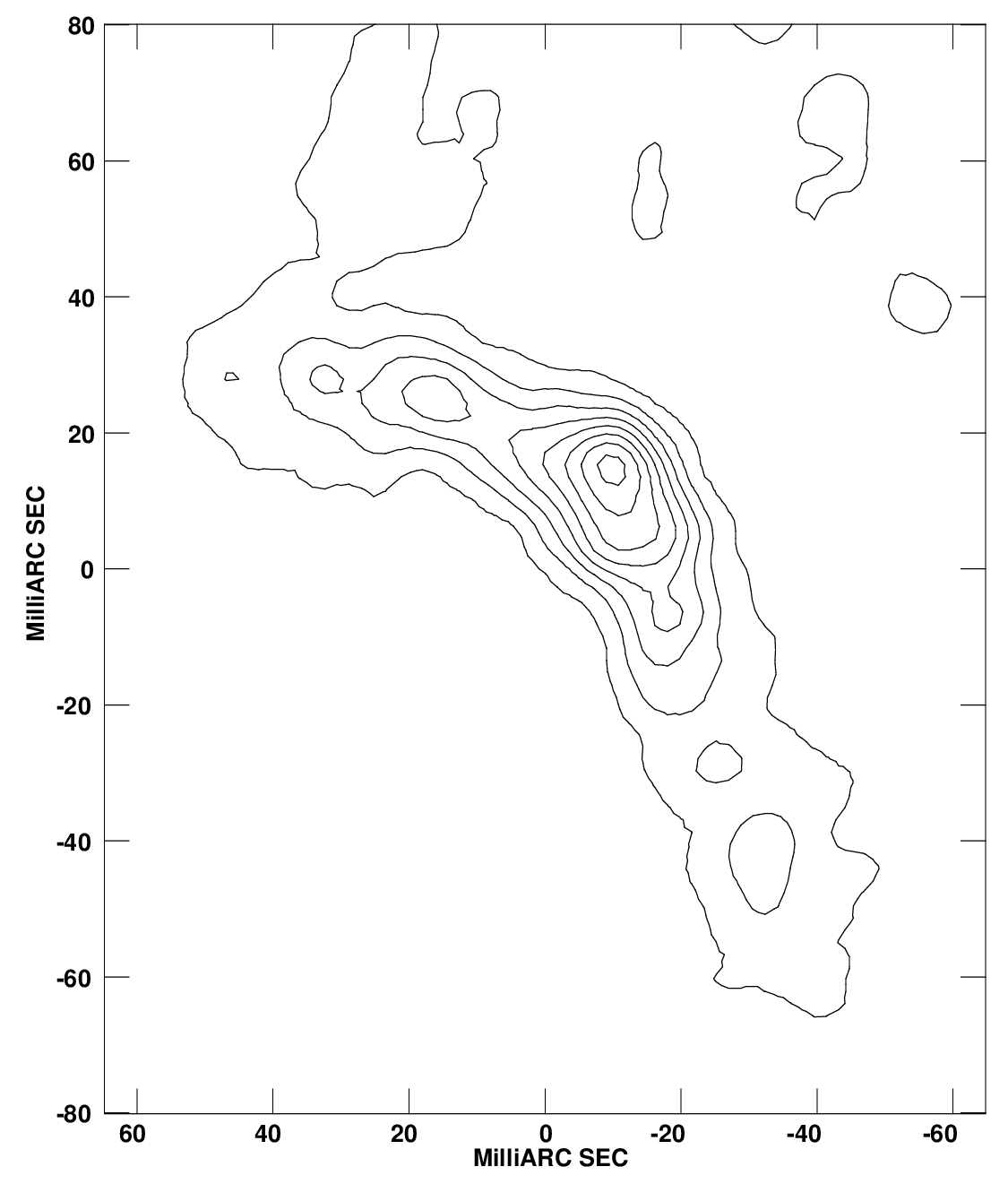}
\caption{{\it a) Left:} VLBA+VLA image of component SE at 1.37 GHz at 10 mas
resolution. The contours are $-1, 1, 2, 3, 4, {\ldots} \times 0.1$ mJy/beam.
The peak surface brightness is 1.35 mJy/beam and the off-source noise is 36
$\mu$Jy/beam.
{\it b) Right:} EVN+MERLIN image of component SE at 5 GHz at 10 mas resolution.
The contours are $-1, 1, 2, 3, 4, {\ldots} \times 0.04$ mJy/beam.
The peak surface brightness is 0.34 mJy/beam and the off-source noise is 15
$\mu$Jy/beam.
}
\end{figure*}

\subsection{Component SE in Mrk 273}
In the past it has been suggested that this is a
background source, but evidence for gas infall
in the HI 21 cm absorption image and the low probability
of a chance projection are against the background source hypothesis
(Carilli \& Taylor \cite{CT00}).
The 1.4 GHz images do not clarify if the source is core jet (and hence AGN
driven) or a compact starburst, and observations
in other bands don't provide a unique interpretation.
At 5 GHz, the SE component is resolved in an arc-shaped radio emission
resembling a core twin-jet source.
The most striking peculiarity of the SE component  is the steep spectral index.
The integrated value is $\alpha\simeq 1.4$ with values ranging from 0.9 to
1.6 across the source. At the full resolution of the EVN observations (5 mas)
the source is completely resolved out confirming the absence of any high
brightness radio feature as already noted by Carilli \& Taylor (\cite{CT00}).

\section{Summary}

We have reduced and analyzed 5 GHz EVN+MERLIN observations of the ULIRG
Mrk~273. Both the sub-arcsecond scale compact nuclei, N and SE, were detected
and resolved. Using published images at 1.4 GHz we have derived the spectral
index of the radio emitting regions reaching the following results:
\begin{enumerate}
\item The compact component N1, often indicated as a possible AGN nucleus,
 has a very steep spectral index ($\alpha\simeq 1.2$). Unless strong flux
density variability has occurred between the two epochs of observation this
result is difficult to reconcile with the AGN hypothesis.
\item The component N2 is partly resolved in several compact radio sources.
The integrated spectral index of this region is flat ($\alpha=0.15$) probably 
because of the superposition of several components with peaked spectra 
and/or free-free absorption. This is consistent with findings in the NIR which
identify N2 as a compact region with the strongest star formation.
\item The spectral index of the extended emission in component N is typical
of non-thermal optically thin radio emission ($\alpha\simeq 0.8$), and
the luminosity of the extended emission is consistent 
with being produced by electrons diffused away from supernova remnants in a
luminous starburst.
\item The SE component has a very steep spectrum ($\alpha\simeq 1.4$), with
no compact high brightness component. 
\end{enumerate}

\begin{acknowledgements}
We thank Chris Carilli for the processed VLBA images at 1.4 GHz of Mrk
273, and the staff at JIVE for the efforts spent during the 
correlation and pipeline of these data.
\end{acknowledgements}

\end{document}